\documentstyle[aps,prd,preprint,tighten,eqsecnum,amssymb,epsf,psfrag]{revtex}
\newcommand{\be}{\begin{equation}}
\newcommand{\ee}{\end{equation}}
\newcommand{\bd}{\begin{displaymath}}
\newcommand{\ed}{\end{displaymath}}
\newcommand{\ba}{\begin{array}}
\newcommand{\ea}{\end{array}}

\newcommand{\beq}{\begin{eqnarray}}
\newcommand{\eeq}{\end{eqnarray}}
\newcommand{\bi}{\begin{itemize}}
\newcommand{\ei}{\end{itemize}}

\newcommand{\rhoS}{\rho_{S}}
\newcommand{\rH}{R_{H}}
\newcommand{\rhoH}{\rho_{H}}
\newcommand{\rhoP}{\rho_{P}}
\newcommand{\rhoE}{\rho_{E}}
\newcommand{\dP}{d_{P}}
\newcommand{\dH}{d_{H}}

\newcommand{\qprimeH}{Q'_{H}}
\begin{document}
\title{SU(2) Cosmological Solitons
      }
\author{
        C. Lechner${}^{1}$,
        S. Husa${}^{2,3}$,
        P. C. Aichelburg${}^{1}$
       }
\address{
${}^{1}$ Institute for Theoretical Physics\\
         University of Vienna, Boltzmanng. 5, A-1090 Wien, Austria \\
${}^{2}$ Department of Physics and Astronomy,
         University of Pittsburgh\\
         Pittsburgh, PA 15260, USA\\
${}^{3}$ Max-Planck-Institute for Gravitational Physics,
         Albert Einstein Institute\\
         Am M\"uhlenberg 1, D-14476 Golm, Germany\\
        }
\date{\today}
\maketitle

\begin{abstract}
We present a class of numerical solutions to the SU(2) nonlinear
$\sigma$-model coupled to the Einstein equations with cosmological constant
$\Lambda\geq 0$ in spherical symmetry.
These solutions are characterized by the presence of a regular static region
which includes a center of symmetry. They are parameterized by a dimensionless
``coupling constant'' $\beta$, the sign of the cosmological constant, and an
integer ``excitation number'' $n$.
The phenomenology we find is compared to the
corresponding solutions found 
for the
Einstein-Yang-Mills (EYM) equations with positive $\Lambda$ (EYM$\Lambda$).
If we choose $\Lambda$ positive and fix $n$, we find a family of static
spacetimes with a Killing horizon for $0 \leq \beta < \beta_{max}$.
As a limiting solution for $\beta = \beta_{max}$ we find a {\em globally}
static spacetime with $\Lambda=0$, the lowest excitation being the
Einstein static universe.
To interpret the physical significance of the Killing horizon in the
cosmological context, we apply the concept of a trapping horizon as formulated
by Hayward. 
For small values of $\beta$ an asymptotically de Sitter dynamic region
contains the static region within a Killing horizon of cosmological type.
For strong coupling the static region contains an ``eternal cosmological
black hole''.
\end{abstract}

\section{Introduction}

The aim of this paper is to discuss the static spherically symmetric
solutions of the SU(2) nonlinear $\sigma$-model coupled to the Einstein
equations with cosmological constant $\Lambda \geq 0$
 (subsequently referred to as the E$\sigma_{SU(2)}\Lambda$--model).
The existence of such solutions was suggested by the fact, that the
Einstein universe \cite{Bizon1}
and the de Sitter spacetime \cite{PCAichCL} admit the existence of a discrete
one-parameter family of static, spherically symmetric regular solutions of
the  $SU(2)$-$\sigma$-model. These solutions are unstable, the number of
unstable modes equals the ``excitation index''.
The static solutions on the de Sitter background correspond to the uncoupled
limit, i.e. vanishing coupling constant of the
E$\sigma_{SU(2)}\Lambda$-model.

Globally regular stationary ``solitonic'' solutions, if they exist, play a
fundamental role in the dynamics of gravitating systems. If they are stable,
they provide possible end states of evolution -- and models for ``stars'' or
``particles''.
Interest in solitonic solutions and the subtleties of the interaction of
matter with gravity was revived by the surprising numerical discovery of
static solutions to the EYM system by Bartnik and McKinnon \cite{BMcK}.
It was soon found that these solutions are unstable \cite{BMcK unstable}, but
there is a link between this instability and another
surprise in GR:

In recent years the discovery of critical phenomena at the threshold of black
hole formation has introduced a new twist into the gravitational collapse
problem. In the original work of Choptuik the intermediate attractor
that separates the two possible generic end-states of a free scalar field
--
the formation of a black hole and complete dispersion
--
is the self-similar ``choptuon'' which forms a naked singularity.
However, as was discovered by Choptuik, Chmaj and Bizon \cite{Bizon2},
also unstable solitons can serve as an intermediate attractor in critical
collapse. In particular they found that the first Bartnik-McKinnon
excitation
actually forms an intermediate attractor associated with type I critical
collapse phenomena --  as opposed to the type II originally found by
Choptuik (both named because of the correspondence to phenomena of
statistical physics, for a recent overview on critical collapse phenomena
see e.g. Choptuik \cite{choptuik-puna}).

While both the free scalar field and the EYM-system do not contain any
dimensionless parameters, the $\sigma$-models are simple systems 
with a dimensionless coupling constant. Thus they provide a convenient
family of theories that is suitable to study genericity and
bifurcation phenomena.

It is known that, apart from trivial solutions, the nonlinear $\sigma$-models
do not admit soliton solutions in Minkowski space, and when coupled to
gravity, there exist neither solitons nor static black hole solutions that
are asymptotically flat (see e.g. \cite{Heusler-1996}). 
The presence of a positive cosmological constant changes this
situation by introducing a length scale into the model. From dimensional
analysis one concludes that the behavior of the solutions depends
nonperturbatively on $\Lambda$, and that only the sign of $\Lambda$
is significant.

Within the static spacetimes we restrict ourselves to spacetimes which
possess a static region that has a regular center and is bounded
by at most one Killing horizon (as opposed to two, such as in the
Schwarzschild-de Sitter case).

In the coupled case we find four qualitatively different types of solutions
along each ``branch'' defined by the ``excitation index''.
For small coupling constant of the $\sigma$-model, the solutions are globally 
regular, and a Killing horizon separates a static region from an
asymptotically de Sitter dynamic region.
For intermediate values of the coupling constant the situation is similar,
but the region exterior to the cosmological horizon expands for some time,
but eventually recollapses.
For even higher values of the coupling constant the horizon encloses 
a dynamic region that undergoes collapse. Thus in some sense the static
region 
gets turned ``inside out''. 

Solutions with a finite regular static region and positive cosmological
constant only exist for a coupling
constant smaller than some critical value which depends on the excitation
index. The limit is singular, but combining it with a limit
$\Lambda \rightarrow 0$ yields solutions which are globally regular and 
static and have spatial topology $S^3$.

For $\Lambda > 0$
the ``first static excited states'' exhibit
one unstable mode, so that our solutions form a one parameter-family of
candidate solutions for type I intermediate attractors.
The full linear stability analysis, that leads to this result,
will be presented in a separate paper \cite{cosmoStoer}.

Similar results have been found for the EYM$\Lambda$ system, which has been
studied in great detail by Volkov et. al. \cite{Brodbeck1}.

The organization of this paper is as follows: In Sec. \ref{sec:sigma-model}
we introduce our nonlinear $\sigma$-model and discuss its basic properties.
We then specialize to spherical symmetry using the
hedgehog ansatz for the $SU(2)$-valued matter field and write out the
spherically symmetric field equations in a suitable gauge.

The problem of finding regular static solutions of the
E$\sigma_{SU(2)}\Lambda$ problem is
discussed in Sec. \ref{sec:static}. The static regions containing the soliton
are confined within a Killing horizon, outside of which the timelike
Killing vector becomes spacelike and  the spacetimes are dynamical.
The static regions are constructed by solving a boundary 
value problem as described in Sec. \ref{subsec:bdryvalprobl}. 
The global structure of the spacetimes is then determined by evolving the
static data beyond the horizon.
This evolution problem is discussed in Sec. \ref{sec:evolution}.
The phenomenology of the solutions we find is presented in
Sec. \ref{sec:phenomSol}. 

Finally, Sec. \ref{sec:discussion} gives a discussion of
our results and compares them to the results of Volkov et. al. 
\cite{Brodbeck1} for the EYM${\Lambda}$-system.

Conventions are chosen as follows:
spacetime indices are greek letters, SU(2) indices are uppercase latin letters,
the spacetime signature is $(-,+,+,+)$, the Ricci tensor is defined
as ${\mathcal R}_{\mu \nu} = {{\mathcal R}_{\mu\lambda\nu}}^{\lambda}$,
and the speed of light is set to unity, $c=1$.

\section{The SU(2)-$\sigma$-model in spherical symmetry}\label{sec:sigma-model}

\subsection{Harmonic Maps as Matter Models}

Nonlinear $\sigma$-models are special cases of harmonic maps from a spacetime
$({\mathbf M},g_{\mu\nu})$ into some target manifold $({\mathbf N},G_{AB})$
(See e.g. \cite{Heuslerbuch}).
For the SU(2)-$\sigma$-model, the target manifold is taken as $S^3$
with $G_{AB}$ the ``round'' metric of constant curvature.
Harmonic maps $X^{A}(x^{\mu})$ are extrema of the simple geometric action
\begin{equation}\label{SM}
S = \gamma\int_{{\mathbf M}} d^{m}x \sqrt{|g|}
g^{\mu\nu}\partial_{\mu}X^{A}\partial_{\nu}X^{B}G_{AB}(X). 
\end{equation}
Their importance for physics was pointed out in a review
article by Misner \cite{Misner}.

Variation of the matter action (\ref{SM}) with respect to the metric 
$g_{\mu \nu}$ yields the stress-energy-tensor of the harmonic map
\be\label{StressEnergy}
T_{\mu\nu} = \gamma ( \nabla_{\mu} X^A \nabla_{\nu} X^B G_{AB}(X^C) -
   \frac{1}{2}g_{\mu\nu} \nabla^{\sigma}X^A \nabla_{\sigma} X^B G_{AB}(X^C) ).
\ee
This stress-energy-tensor obeys the weak, strong and dominant energy
conditions. 

In order to produce static configurations attractive and repulsive forces
have to be balanced. In particular in flat space the virial theorem implies
that the components of the stress-energy tensor cannot have a fixed sign
(see the discussion by Gibbons \cite{Gibbons}).
For static configurations of any harmonic map the sum of the principal
pressures  $\sum_{{\hat i}=1}^{3} T_{{\hat i}{\hat i}}$ is nonpositive
everywhere \cite{Gibbons} 
($T_{{\hat i}{\hat j}}$ denote the spatial components of the 
stress-energy tensor with respect to an orthonormal frame).
The self-interaction of any $\sigma$-model can therefore be interpreted as
attractive.
In contrast for the Yang Mills field the sum of the principal pressures is
nonnegative, which can be interpreted as a repulsive self-interaction.
Both fields therefore do not allow soliton solutions on a flat background.
But while the Yang-Mills field coupled to gravity admits solitons 
by cancelling the repulsive 
force with gravity \cite{VolkovGaltsov},
the gravitating $\sigma$-model correspondingly does not (see e.g. 
\cite{Heusler-1996}).

Thus we cannot expect static solutions to exist unless we add
some ``repulsive force'' - or in the presence of nontrivial topology.
As an example for the latter one could think of a static spherical
universe of topology $S^3$, where there is a balance between the tendencies
to collapse towards either ``center''. 
A repulsive force on the other hand can be introduced with a positive
cosmological constant $\Lambda$.
Therefore we consider the total action
\be\label{Action}
S = \int d^4 x \sqrt{-g}(\frac{1}{16 \pi G}({\mathcal R} 
- 2 \Lambda ) + {\mathcal L}_{M}),
\ee
where ${\mathcal L}_{M}$ is given by the Lagrangian of (\ref{SM}).
This action gives rise to the Einstein equations
\be\label{EinsteinEqus}
{\mathcal R}_{\mu\nu} - \frac{1}{2} g_{\mu\nu} {\mathcal R} + 
\Lambda g_{\mu\nu} = 8 \pi G T_{\mu\nu},
\ee
as well as to the field equations of the harmonic map 
\be\label{FieldEqusHM}
g^{\mu\nu}(\nabla_{\mu}\nabla_{\nu} X^A + 
     {\tilde \Gamma^A_{BC} (X^D) \nabla_{\mu}X^B \nabla_{\nu} X^C}) = 0,
\ee
where ${\tilde \Gamma^A_{BC}}$ denote the Christoffel symbols with respect to 
$G_{AB}$. The field equations thus allow a general nonlinear dependence on
the fields $X^A$ through the Christoffel symbols and they depend quadratically
on the first derivatives of the fields. 

In units where $c = 1$ the coupling constant $\gamma$ of the harmonic map
has dimension $mass/length$, whereas the gravitational constant $G$ is of
dimension $length/mass$. Both constants enter the equations only in the
dimensionless product $\beta = 4 \pi G \gamma$, thereby defining a
one-parameter family of distinct gravitating matter models.
The parameter $\Lambda$ plays a different role, since it has dimension
$1/length^2$.
Thus, when the cosmological constant is nonzero, it provides the
length scale of these theories.
Therefore all theories with the same value of $\beta$ are equivalent
irrespective of the value of $\Lambda$.
If, on the other hand, $\Lambda=0$, the field equations are scale invariant.
The gravitating SU(2)-$\sigma$-model with non-negative cosmological constant
thus corresponds to a two-parameter family of inequivalent theories,
parameterized by the continuous parameter $\beta$ and the discrete
parameter sign$(\Lambda)$.

\subsection{Spherical Symmetry}\label{subsec:sphersymm}

A spherically symmetric metric can be written in the general form as the
warped product
(see e.g. \cite{Hawking-Ellis_AppendixB})
\be\label{genSSMetric_short}
ds^{2} = d\tau^{2} + R^{2}(\rho,t) 
d\Omega^{2},
\ee
where $d\tau^{2}$ is a general two-dimensional Lorentzian line-element
and the area of the orbits of SO(3) is given by $4 \pi R^2(\rho,t)$.
Choosing the coordinates $t$ and $\rho$ orthogonal one can write the metric 
in the form  
\be\label{genSSMetric}
ds^{2} = - A(\rho,t) dt^{2}
         + B(\rho,t) d \rho^{2} 
         + R^{2}(\rho,t) d\Omega^{2}.
\ee
In (\ref{genSSMetric}) there is some gauge freedom left, which can be used 
to eliminate one of the three functions $A$, $B$ or $R$.

One way to fix the gauge would be to use the function $R$ as a 
coordinate. This is possible as long as $\nabla_{\mu} R \ne 0$. 
This gauge is usually referred to as the Schwarzschild gauge.
Here we will deal with the situation that $\nabla_{\mu} R$ becomes zero 
on some maximal two-sphere (which is not a horizon), a phenomenon which
we will discuss in detail below. 

Following  
Volkov et al. \cite{Brodbeck1} 
we therefore choose a different gauge, which keeps close to the standard
Schwarzschild line element in another way: we fix $A(\rho,t) = 1/B(\rho,t)
\equiv Q(\rho,t)$ and keep the area of the SO(3) orbits as the second free
function:          
\be\label{ourSSMetric}
ds^{2} = - Q(\rho,t) dt^{2} + \frac{1}{Q(\rho,t)} d \rho^{2} 
         + R^{2}(\rho,t) d\Omega^{2}.
\ee
The metric (\ref{ourSSMetric}) is 
regular where $Q(\rho,t)$ and $ R(\rho,t)$ are
regular functions on spacetime, except possibly at points where 
either $Q(\rho,t)$ or $R(\rho,t)$ vanish.
The vanishing of $Q(\rho,t)$ is related to the existence of horizons,
which will also be discussed below.

For a regular spacetime the vanishing of $R(\rho,t)$ is associated with 
the singularity of spherical coordinates at the axis of symmetry.
Such a coordinate center
does not necessarily have to be present (e.g. in a spherical wormhole),
but we will require our spacetimes to possess at least one regular 
center -- there may also be two, such as in the 
Einstein static universe written in 
spherical coordinates.
In the following, several boundary conditions will be derived from regularity
requirements, we will therefore make some comments on regularity 
near the center of spherical symmetry
for the metric defined in Eq. (\ref{ourSSMetric}) :
We assume the existence of four regular (meaning $C^{\infty}$ or 
$C^{k}$ as appropriate) coordinate
functions $x$, $y$, $z$, $t$ on the manifold. All other functions are defined 
as regular
if they can be expressed as regular functions of  $x$, $y$, $z$, $t$.
Furthermore, the relations between the functions
$R$, $\theta$, $\phi$  and  $x$, $y$, $z$
are required to be the standard coordinate transformation between Cartesian
and spherical coordinates:
$x = R \sin\phi \sin\theta$,
$y = R \cos\phi \sin\theta$,
$z = R          \cos\theta$.
Note that the spherical coordinates, in particular $R$, are not 
regular functions,
but all even powers of $R$ are. Near the axis we choose the
parameterization of the radial coordinate $\rho$ such that it has the same
regularity features as $R$:
 $$\rho = R \, h(R^2),$$
where $h$ is a regular positive function --  and thus also $\rho$ itself is 
not regular.
Any spherically symmetric function which is a regular function of
$x$, $y$, $z$, $t$ (or just ``regular'' for short) can therefore be written as
a regular function of $\rho^2$ and $t$.

For any fixed $t$ the center $R=0$ then is a regular point of spacetime, if
$Q(0,t) = h(0)$. Without restricting generality we can choose
$h(0)=1$ for convenience, such that
\begin{eqnarray}\label{metricO}
R(\rho,t) &=& \rho (1 + O(\rho^2)) \nonumber\\
Q(\rho,t) &=&       1 + O(\rho^2).
\end{eqnarray}

In spherical symmetry it is possible to define a quasilocal mass 
\be\label{def:QLM_general}
m = \frac{R}{2}(1 - \nabla_{\mu}R \nabla^{\mu} R),
\ee
which reads
\be\label{QLM}
m = \frac{R}{2} (1 - \frac{{\dot R}^2}{Q} - Q (R')^2).
\ee
in our coordinate system (\ref{ourSSMetric}). 
For a recent discussion of the properties of the quasilocal mass in spherical
symmetry see Hayward \cite{Hayward-QLM}.

The harmonic map field configuration $X^{A}(x^{\mu})$ can be called 
spherically symmetric,
if the Lie derivatives of the energy-momentum tensor with respect to the 
Killing vector
fields that generate the SO(3) action vanish. 
One possibility to achieve this, is to demand that all fields be functions 
of $\rho$ and $t$ only. This would leave us with a coupled system of
differential equations for three fields and two metric functions. 
Since the target manifold $(S^3,G)$ {\em also} admits SO(3) as an isometry
group, there is an alternative way to impose spherical symmetry on the
harmonic map, which is the well known hedgehog ansatz. We first introduce
spherical coordinates $(f, \Theta, \Phi)$ on the target manifold,
writing the SU(2) line element as
\be
ds^2 = df^2 + \sin^2 f (d\Theta^2 + \sin^2\Theta d\Phi^2).
\ee    
The hedgehog ansatz now ties the coordinates on the target manifold
to those on the base manifold:
\be    
f(x^{\mu}) = f(\rho,t), \quad \Theta(x^{\mu}) = \theta, \quad 
      \Phi(x^{\mu})=\varphi.
\ee
Due to this ansatz two of the three coupled fields are already determined
and only one field $f(\rho,t)$ enters the equations.
The matter field equations (\ref{FieldEqusHM}) are then reduced
to the single nonlinear wave equation
\be\label{eq:WaveEq}
\Box f = \frac{\sin(2 f)}{R^2},
\ee
where $\Box$ is the wave operator of the spacetime metric. 

Regularity of the harmonic map within this ansatz means that when expressing
the field in terms of regular coordinates $X^A$ on SU(2), the $X^A$ are
regular fields on spacetime. Since $X^A(x^\mu) = (f(\rho,t)/\rho) x^A$,
where $x^A$ denote the Cartesian coordinates $x,y,z$ on spacetime,  
the field $f(\rho,t)$ thus has to be of the form
\be\label{fieldO}
f(\rho,t) = \rho \, H(\rho^2)  = f'(0) \ \rho + O(\rho^3), 
\ee
where $H$ is a regular function. Thus $f$ vanishes at the center, which means
that the center $R=0$ is mapped to one of
the poles of the target manifold, that have been fixed via the hedgehog
ansatz.

\section{Static Solutions}\label{sec:static}

\subsection{Static Field Equations}\label{subsec:staticEqs}

In addition to spherical symmetry we assume that spacetime admits a
hypersurface orthogonal Killing vector field $\partial_t$, which is timelike
in some neighborhood of the (regular) center $R=0$. 

The combinations $({}^t_t) - ({}^{\rho}_{\rho}) - 2 ({}^{\theta}_{\theta})$ and
$({}^t_t) - ({}^{\rho}_{\rho})$ of the mixed components of Einstein's 
equations plus the matter
field equation  (\ref{eq:WaveEq}) then yield the following
set of coupled second order autonomous ODEs for the metric
functions $Q(\rho), R(\rho)$ and the matter field $f(\rho)$: 
\beq\label{staticequ2}
(R^{2} Q')' & = & - 2 \Lambda R^{2},\label{Q}\\
R'' & = & - \beta R f'^{2},\label{R}\\
(Q R^{2} f')' & = & \sin 2f.\label{f}
\eeq
Furthermore the above system of equations 
has a constant of motion:
\be\label{constofm}
2 \beta \sin^{2}f + R^{2}(\Lambda - \beta Q f'^{2}) + R Q' R' + Q R'^{2} 
                                     - 1 = 0.
\ee

This expression can either be derived by integrating the system 
(\ref{Q})-(\ref{f}) and using the regularity conditions at the 
axis, or by using the $({}^{\rho}_{\rho})$ component of
Einstein's equations.

Note that if $f$ is a matter field solution, then so is also
$k\,\pi \pm f$ for any integer $k$.

If $\Lambda$ is nonzero it sets the length scale, it thus can be eliminated 
in these equations by using the dimensionless quantities
${\bar \rho} = \sqrt{\Lambda} \rho$ and ${\bar R} = \sqrt{\Lambda} R$.
If, on the other hand, $\Lambda$ vanishes, the equations are scale invariant,
and thus invariant under rescalings ${\bar \rho} = a \rho$ and ${\bar R}=a R$.
Furthermore, in this case Eq. (\ref{Q}) together with regularity conditions at
the axis implies $Q\equiv 1$.

If we integrate the equations with the above boundary conditions 
from $\rho=0$ to larger values of $\rho$ one of the following four
situations has to occur: 
\begin{enumerate} 
\item the static region ``ends'' in a singularity,
\item integration might run into a second (regular) pole $R=0$, which would
mean that the resulting spacetime has compact slices of constant $t$ and
is  globally static,
\item the static region might persist up to spatial infinity, 
\item the static region might be surrounded by a Killing horizon beyond 
    which spacetime becomes dynamical.
\end {enumerate}
The first case can easily be produced by shooting off a regular center with
arbitrary initial data, it can be excluded by setting up a boundary value
problem that enforces one of the other three cases.
The second case can be discarded for our model with positive cosmological
constant, since the existence of a static region with two regular
centers is not compatible with the field equations:
To see this recast Eq. (\ref{Q}) into the integral form 
(\ref{Qprime}). It is clear then, that - for nonzero $\Lambda$ -   
$Q'$ diverges if $R$ goes to zero a second time. For $\Lambda = 0$ of
course, such solutions may exist as e.g. the static Einstein universe
(\ref{EinsteinU}), which will be discussed in the next section.  
The third case can also be excluded if the cosmological constant is
positive: a static region can not be extended to spatial infinity, but rather
a singular point of the equations $Q(\rhoH)=0$, 
that corresponds to a horizon, has to develop.
This can be seen as follows: 
Suppose the solution exists up to $\rho \to \infty$ and $R$ is monotonically
increasing - all other assumptions automatically lead to one of the cases
1, 2 or 4 - then 
one can show, that for $\Lambda$ positive 
$Q' < - const. /\rho$ for $\rho$ large enough, 
which means that $Q$ would be 
bounded from above by a function that tends to $-\infty$ as 
$\rho \to \infty$. So again $Q$ has to cross zero at some finite 
value of $\rho$.
    
For positive cosmological constant we therefore may confine ourselves
to the cases with horizon, and we will construct static regions as an ODE
boundary value problem, where the boundary conditions correspond
to a regular center at $\rho=0$ and a regular horizon at $\rho=\rho_{H}$,
where the determination of the value of $\rho_{H}$ is part of the boundary
value problem. The appropriate boundary conditions at the horizon
will be determined in Sec. \ref{subsec:horizons}, and the boundary value 
problem is described in Sec. \ref{subsec:bdryvalprobl}.

Given a static region, i.e. a region of spacetime where the Killing vector
$\partial_t$ is timelike, which is bounded by a Killing horizon,
we can consider as a second step the time evolution problem of 
the static data on the horizon into the dynamic region where the Killing vector
$\partial_t$ is spacelike and the spacetime is thus homogeneous. The
time evolution problem thus reduces to a system of ODEs which are solved as
an initial value problem as described in Sec. \ref{sec:evolution}.

\subsection{Exact Solutions}\label{subsec:exactsols}

Some solutions of Eqs. (\ref{Q})-(\ref{constofm}) with a regular center
can be given in closed form. They do arise as certain limits of the numerically
constructed family of solutions to be given below.

First of all, for 
$\Lambda > 0$, for the trivial case $f \equiv 0$ Eq. (\ref{R}) has the
solutions $R(\rho) = a \rho + b$. Imposing the regularity conditions
(\ref{metricO})
this gives de Sitter spacetime 
\be\label{deSitter}
R(\rho) = \rho, \quad Q(\rho) = 1 - \frac{\Lambda \rho^2}{3}, \quad
f \equiv 0.
\ee 

For $f \equiv \pi /2$, $R(\rho)$ is of the same form as above. For $b=0$ we
get
\be\label{singularsol}
R(\rho) = \rho, \quad Q(\rho) = 1 - 2 \beta - \frac{\Lambda \rho^2}{3},
\quad f \equiv \frac{\pi}{2},
\ee 
which looks like de Sitter for large $\rho$ but has a conical singularity
at the center if $\beta > 0$. 
Furthermore the static region shrinks with increasing $\beta$ and ceases to
exist for $\beta = 1/2$. In the limit of vanishing coupling constant
$\beta=0$, where spacetime is de Sitter, this solution $f \equiv \pi/2$
is still singular at the center with diverging energy density.
Nevertheless the total energy is finite, and this solution may be viewed
as the ``high excitation'' limit of the regular solutions, that exist
on de Sitter background \cite{PCAichCL}.

Finally, for $\Lambda = 0$ Eq. (\ref{Q}), together with 
regularity conditions at the axis, yields $Q \equiv 1$. 
For $\beta = 1$ the remaining equations can be solved analytically to give
the static Einstein universe: 
\be\label{EinsteinU}
R(\rho) = \sin \rho, \quad  Q(\rho) \equiv 1, \quad f(\rho) = \rho.
\ee   
Note that the stress-energy tensor, which has the form of a perfect fluid
in this case satisfies $\mu + 3 p =0$.
As will be described in detail in Sec. \ref{subsec:staticLambdazero} 
the static Einstein universe arises in the limit of maximal coupling 
constant of the numerical constructed first excitation 
iff at the same time $\Lambda$ is set to zero.

For completeness we mention that there also exist exact solutions to
Eqs. (\ref{Q})--(\ref{constofm}), that do not posses a regular center of
spherical symmetry, such as the Nariai spacetime.

\subsection{Horizons and global structure}
      \label{subsec:horizons}

In order to discuss the global structure of the spacetime and in particular
regularity questions from which we derive the boundary conditions for
our ODEs, it is helpful to consider a coordinate system, which is regular
at the horizon. We write the metric (\ref{ourSSMetric}) as
\be\label{def:null_metric}
ds^2 = - Q(\rho) du^2 - 2 du \, d\rho + R(\rho)^2 d \Omega^2,
\ee
where the coordinate function $\rho$ and the metric function $Q(\rho)$
coincide with those in the metric (\ref{ourSSMetric}), 
and the coordinate $u$ is given as
\be
u = t - \int \frac{d\rho}{Q(\rho)}.
\ee

Note that the coordinates (\ref{def:null_metric}) 
cover only half of the maximally extended spacetime. In the following,
we will simplify our discussion by only talking about the Killing horizon
contained in the portion of spacetime covered here.
All statements made can be extended trivially to the complete spacetime
and in particular the second component of the horizon by time reflection.
We also remark that all solutions have the topology $S^3 \times R$.

The static Killing vector field is 
$\partial/\partial u = \partial/\partial t$,
where the latter is taken with respect to the $(t,\rho)$ coordinates.
The metric (\ref{def:null_metric}) is 
regular if $Q(\rho)$ and
$R(\rho)$ are regular functions, except when $R=0$, which corresponds
either to the usual
coordinate singularity of spherical symmetry, which has been discussed
in Sec. (\ref{subsec:sphersymm}) or to a spacetime singularity, as discussed
in Sec. (\ref{sec:evolution}).

The Killing vector field $\partial_u$, which we have assumed to be timelike
in some neighborhood of the (regular) center $R=0$ need not necessarily be
{\em globally} timelike. Regions where it becomes spacelike, i.e. $Q(\rho)<0$,
are dynamic with homogeneous spacelike slices of constant time $\rho$, the
spatial topology being $S^2 \times R$.
Such regions thus correspond to Kantowski-Sachs models.
At the boundary of static and dynamic regions the metric function
$Q(\rho)$ vanishes, such a surface of $\rho=const.$ is thus a null surface.
Furthermore the  Killing vector field $\partial_u$ is null and tangent 
on this surface, which therefore is a Killing horizon.

In the presence of an asymptotic infinity, such as in an asymptotically flat
or asymptotically de Sitter spacetime, one can use the asymptotic region
to classify event horizons, e.g. as black hole or cosmological event horizons.
Furthermore one is provided a straightforward definition of ``inward'' and
``outward'' directions.
In the cosmological case these issues are less clear.
We therefore follow the definitions of Hayward
\cite{Hayward-trapping_horizon}:
He proposed a local definition of a {\em trapping horizon}, a concept which
does not make use of asymptotic flatness and is therefore also suitable for
more general situations.
Intuitively, the physical interpretation of the Killing horizon depends on
whether the dynamical region is collapsing or expanding off the horizon, 
and whether this region is to be interpreted as {\em inside} or {\em outside}.
Both can be formulated in terms of the null expansions
$\Theta_{\pm}$ of ingoing and outgoing null rays from two-surfaces. In
spherical symmetry we consider $R=const.$ surfaces, with null expansions
\be\label{thetaplusminus}
\Theta_{\pm} = \frac{1}{R^2} {\mathcal L}_{\pm} R^2,
\ee
where ${\mathcal L}_{\pm}$ is the Lie-derivative along the null directions
\be
l_+ = \partial_{\rho} \qquad \textrm{and} \qquad l_-= 2 \partial_u -
      Q\partial_{\rho}
\ee
respectively, so 
\be\label{Theta_pm}
\Theta_+ = 2 \frac{R'}{R} \qquad \textrm{and} \qquad 
\Theta_- = - 2 Q \frac{R'}{R}.
\ee
The first use of the expansions is to define a {\em trapped surface} 
in the sense
of Penrose \cite{Penrose} as a compact spatial surface for which
$\Theta_- \Theta_+ > 0$. If one of the expansions vanishes, 
the surface is called a {\em marginal surface}.
For a non-trapped surface $\Theta_-$ and $\Theta_+$ have opposite signs,
and we call directions in which the expansion is positive {\em outward},
and {\em inward} when it is negative.

On the Killing horizon $Q=0$, $\Theta_- |_{Q=0} = 0$ while 
$\Theta_+ |_{Q=0} \ne 0$ and ${\mathcal L}_+ \Theta_- |_{Q=0} \ne 0$
(except when also $R'=0$, which we exclude for the moment).
In Hayward's terminology \cite{Hayward-trapping_horizon} such a 
three-surface is called
a {\em trapping horizon}.
It is said to be {\em outer}  if  ${\mathcal L}_+ \Theta_- |_{Q=0} < 0$,
                 {\em inner}  if  ${\mathcal L}_+ \Theta_- |_{Q=0} < 0$,
                 {\em future} if  $\Theta_+ |_{Q=0} < 0$ and
                 {\em past}   if  $\Theta_+ |_{Q=0} > 0$.
A future outer trapping horizon provides a general definition of a black hole,
while a white hole has a past outer horizon, and cosmological horizons are
inner horizons. 

If  $R'(\rho) = 0$ within the static region then
\be\label{futureouter}
{\mathcal L}_{+} \Theta_{-} |_{Q=0} < 0 , \qquad \Theta_+ |_{Q=0} < 0 
\ee
and the surface $Q=0$ is a {\em future outer} trapping horizon. On the other
hand if $R'$ vanishes in the dynamical region or is monotonic, then the
signs in (\ref{futureouter}) are reversed and one speaks of a {\em past
inner} trapping horizon.

Note, that at points where $R'=0$, both $\Theta_{\pm}=0$. However, this does
not lead to a (trapping) horizon since the expansions merely change sign. 
The significance of such marginal surfaces is that
the meaning of {\em inward} and {\em outward} directions are reversed. A
typical example is an ``equatorial'' two-surface of a round three-sphere
at a moment of time symmetry.

Using the expansions $\Theta_+$ and $\Theta_{-}$ (\ref{Theta_pm}) one can
rewrite the quasilocal mass (\ref{QLM})
\be
m = \frac{R}{2} (1 + \frac{R^2}{4} \Theta_{+}\Theta_{-}).
\ee
Using the above assignment of inward and outward direction the quasilocal
mass can be interpreted as the total mass that is contained {\em within} 
any spatial three volume, that is bounded by the two-sphere with areal
radius $R$. On the marginal surfaces, where $R'=0$, the quasilocal mass
equals $R/2$.

\subsection{The boundary value problem for the static
region}\label{subsec:bdryvalprobl}

The system of ODEs (\ref{Q})--(\ref{f}) has singular points for $R=0$ and
$Q=0$. Making use of the imposed regularity conditions at the axis 
(\ref{metricO}) and (\ref{fieldO}) the equations demand in addition
$Q''(0) = - 2 \Lambda /3$. Note that solutions, that are regular in a
neighborhood of $\rho=0$ are determined by the value of the {\em single}
parameter $c \equiv f'(0)$.

The other singular point occurs at the horizon, where $Q(\rhoH) =0$. 
Formal Taylor series expansions around $\rho = \rhoH$ give
\beq\label{boundaryH}
Q(\rho) & = &\qprimeH (\rho - \rhoH) + \frac{Q''(\rhoH)}{2} (\rho - \rhoH)^{2}
          + O((\rho - \rhoH)^{3}) \nonumber\\
R(\rho) & = & \rH + R'(\rhoH) (\rho - \rhoH) + \frac{R''(\rhoH)}{2}
            (\rho - \rhoH)^{2} + O ((\rho - \rhoH)^{3}) \nonumber\\
f(\rho) & = & b + f'(\rhoH) (\rho - \rhoH) +  \frac{f''(\rhoH)}{2}
            (\rho - \rhoH)^{2} + O ((\rho - \rhoH)^{3}),
\eeq
where 
\be 
\rhoH, \quad f(\rhoH) = b, \quad  R(\rhoH) = \rH, \quad Q'(\rhoH) = \qprimeH
\ee
are free shooting parameters. The other coefficients are determined 
by the requirement that $Q$ and $R$ be regular functions of $\rho$.
Consistency with Eqs. (\ref{Q}) - (\ref{constofm}) then amounts to
the conditions:
\beq
R'(\rhoH)  & = &  \frac{1 - 2\beta \sin^{2}b - \rH^{2} \Lambda}{\rH
\qprimeH},\\
Q''(\rhoH) & = & -2 \frac{R'(\rhoH) Q'(\rhoH)}{\rH} - 2 \Lambda,\\
f'(\rhoH)  & = & \frac{\sin 2b}{Q'(\rhoH)\rH^{2}},\\
f''(\rhoH) & = &  \frac{\sin 2b}{Q'(\rhoH)\rH^{2}}
                  (- \frac{R'(\rhoH)}{\rH} + \cos 2b + 
                   \frac{\Lambda}{Q'(\rhoH)}).
\eeq
In order to determine the spectrum of static solutions for a fixed
coupling constant $\beta$, 
we solve the boundary value problem Eqs. (\ref{Q}) -- (\ref{f})  with
boundary conditions (\ref{metricO}), (\ref{fieldO}) and (\ref{boundaryH})
between axis and horizon. We use a standard two point shooting
and matching method (routine d02agf of the NAG library \cite{NAG}),
where the parameters $f'(0), \rhoH, f(\rhoH), \rH$ and $\qprimeH$ serve as
shooting parameters.

For $\beta = 0$ a discrete one-parameter family of solutions has
already been discussed in \cite{PCAichCL}.
In order to get good initial guesses for the shooting parameters 
for $\beta > 0$, we follow one solution
from $\beta =0 $ up to higher values of $\beta$, interpolating
the values of
the shooting parameters at the present and last ``$\beta$-step'' to obtain
values for the next ``$\beta$-step''.

\subsection{Integration Through the Horizon}\label{sec:evolution}

Since the singularity at the horizon is merely a coordinate singularity, 
we can extend spacetime through the horizon and reintroduce the coordinates
in (\ref{ourSSMetric}) in the dynamic region beyond the horizon.
The Killing vector field $\partial_{t}$ becomes spacelike,
and instead of $t$ the timelike coordinate is $\rho$.
The coupled system of ODEs (\ref{Q}) -- (\ref{f}) together with 
initial conditions (\ref{boundaryH}) therefore constitute an initial
value problem.

While integrating forward in time $\rho$, essentially two
things can happen according to the behavior of $R(\rho)$:
\begin{enumerate}
\item $R(\rho)$ is monotonically increasing for all $\rho > 0$. Then 
time evolution exists for all $\rho > 0$. This can be seen by turning
Eqs. (\ref{Q})- (\ref{f}) into integral equations:
\beq
Q' &=& -\frac{2 \Lambda}{R^{2}} \int\limits_{0}^{\rho} R^{2}d {\bar \rho}
                    \label{Qprime}\\
R' & = & 1 - \beta \int\limits_{0}^{\rho} R f'^{2} d {\bar \rho}
                    \label{Rprime}\\
f' & = & \frac{1}{Q R^{2}} \int\limits_{0}^{\rho} \sin (2 f) d{\bar \rho}.
                    \label{fprime}
\eeq
All first derivatives are bounded as long as $Q$ and $R$ don't go to zero.
Beyond the horizon $Q$ cannot go to zero, since $Q' < 0 $ for all $\rho > 0$ 
and $Q(\rhoH) = 0$. $R$ cannot go to zero since it is monotonically increasing
by assumption and $R(0)= 0$. Therefore $Q,R$ and $f$ stay finite for all 
finite $\rho$.  

From Eq. (\ref{Rprime}) it follows that $0 < R' \le 1$ for all $\rho \ge 0$, so
$R = O(\rho)$ for $\rho \to \infty$. From Eq. (\ref{Qprime}) we get, that
$Q = O(\rho^{2})$ and from Eq. (\ref{fprime}) 
we see that $f'$ goes to zero as $\rho^{-3}$ and therefore $f$ goes 
to a constant at infinity.

\item $R$ develops an extremum at some finite $\rho_{extr} > 0$. Since
$R'' \le 0$ for all $\rho$ and $R'(\rho) < 0$ for all 
$\rho > \rho_{extr}$, which follows from Eq. (\ref{Rprime}),
$R$ goes to zero  at some finite $\rhoS > \rhoH$.

Here we already excluded
the case $\rhoS < \rhoH$ in Sec. \ref{subsec:staticEqs},
since this means a singularity in the static region.

For the same reasons as above, time evolution exists for all $\rho < \rhoS$.
In the limit $\rho \to \rhoS$ a spacetime singularity occurs. This follows
from inspection of the Kretschmann invariant:
\beq\label{kretschmann}
R^{\mu\nu\sigma\tau} R_{\mu\nu\sigma\tau} & = & \frac{1}{R^4} 
   \biggl( 4 + 4 R^2 Q'^2 R'^2 +
   R^4 Q''^2 + \nonumber\\ 
   & + & 8 Q R' \left( R^2 Q' R'' -
   R' \right) + 4 Q^2 \left( R'^4 + 2 R^2 R''^2
   \right) 
   \biggr).
\eeq
Since by assumption $Q(\rho), Q'(\rho)$ and $R'(\rho)$ are negative 
near $\rhoS$ all terms of Eq. (\ref{kretschmann}) are non-negative and at least 
some of them clearly have a nonzero numerator while the denominator vanishes
with some power of $R$.
\end{enumerate}

The construction of solutions beyond the horizon consists of 
solving the initial value problem, i.e. we integrate  
Eqs. (\ref{Q})-(\ref{f}) with initial
conditions (\ref{boundaryH}) for $\rho > \rhoH$, 
where the parameters $\rhoH, f(\rhoH), \rH$
and $\qprimeH$ are determined by the solutions of the boundary value problem
described in Sec. \ref{subsec:bdryvalprobl}.
For numerical integration we used routine d02cbf of the 
NAG library \cite{NAG}.

\section{Phenomenology of Solutions}\label{sec:phenomSol}

\subsection{Phenomenology of Numerically Constructed Solutions with $\Lambda
    > 0$}

For $\beta=0$ the E$\sigma_{SU(2)}\Lambda$ equations decouple into 
Einstein's vacuum equations with $\Lambda$ and the matter field equation 
(\ref{eq:WaveEq}) on the fixed background. With our regularity conditions at
the axis the solution to Einstein's equations is de Sitter space
\be
Q(\rho) = 1 - \frac{\Lambda \rho^{2}}{3}, \quad R(\rho) = \rho,
\ee
whereas the field equation (\ref{f}) admits a discrete one parameter
family of regular solutions \cite{PCAichCL}. Within the static region
these solutions oscillate around $\pi/2$ where energy increases
with the number of the oscillations and is limited from above by the energy 
of the ``singular'' solution $f \equiv \pi/2$.
Outside the cosmological horizon at $\rho = 1$ the solutions 
remain finite and tend to a constant near infinity. 

Increasing the coupling constant $\beta$, the numerical analysis
shows, that solutions of this type
persist as long as $\beta$ does not get too large. 
The qualitative behavior of the field $f$ in the static
region is the same as in the uncoupled case, i.e. the $n$-th excitation
oscillates $n$ times around $\pi/2$, whereas the behavior of the field 
in the dynamic region as well as the behavior of the 
geometry
depend strongly on the value of the coupling constant $\beta$.

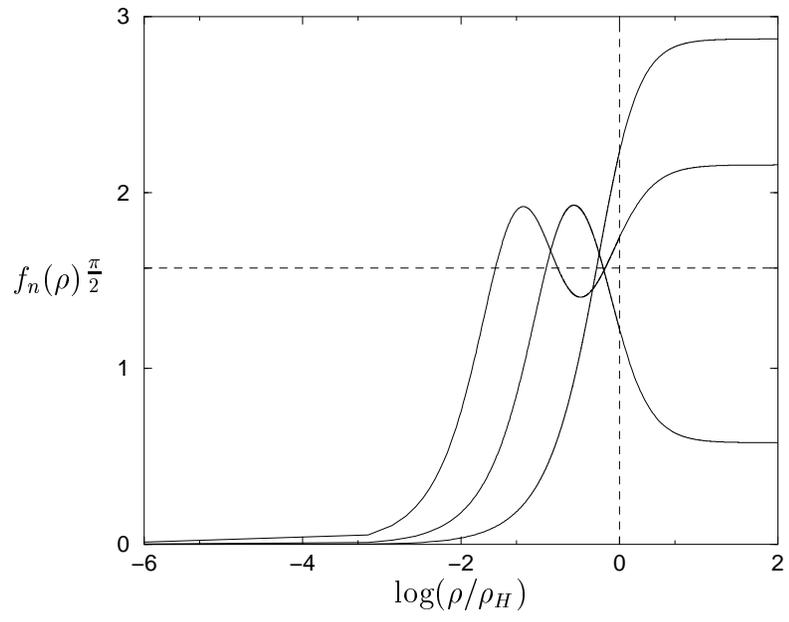
\begin{figure}[h]
\begin{center}
\end{center}
\caption{The first three solutions for coupling constant 
$\beta = 0.3 < \beta_{crit}$. Inside the horizon at $\rho/\rhoH = 1$ the 
``$n$-th'' excitation crosses the $\pi/2$-line $n$ times. 
In the dynamic region outside
the horizon the solutions evolve towards a constant.}
\label{fig:fn_rho_rhoH}
\end{figure}

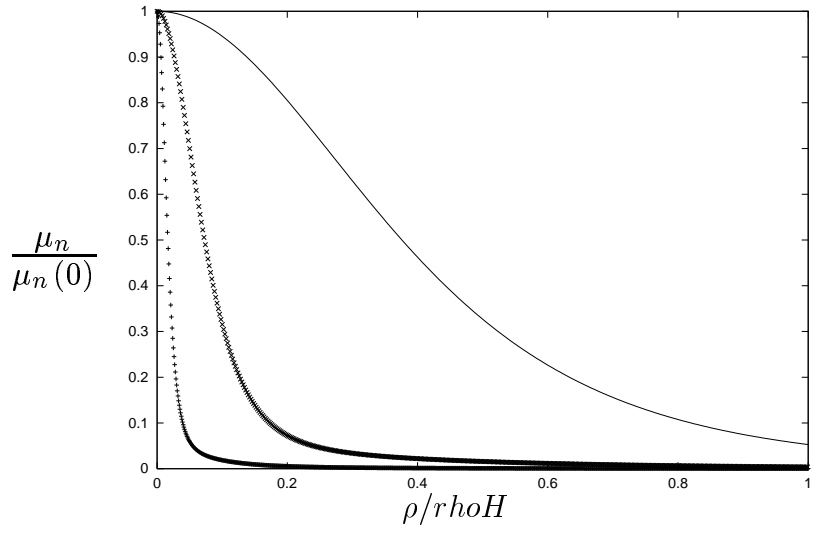
\begin{figure}
\begin{center}
\end{center}
\caption{The energy density $\mu_{n} = Q_{n} (f'_{n})^2 + \sin^2(f_n)/R_n^2$ 
for the first three excitations at $\beta=0.3$.
-- denotes the frist, x the second and + the third excitation, all 
$\mu_n$ have been normalized to unity at the axis. The
corresponding total energies, measured in units of $1/\sqrt{\Lambda}$ are 
$E_1/4\pi\gamma= 1.66621, E_2/4\pi\gamma = 1.88082$ and  
$E_3/4\pi\gamma = 1.956613$.}
\label{fig:endens}
\end{figure}

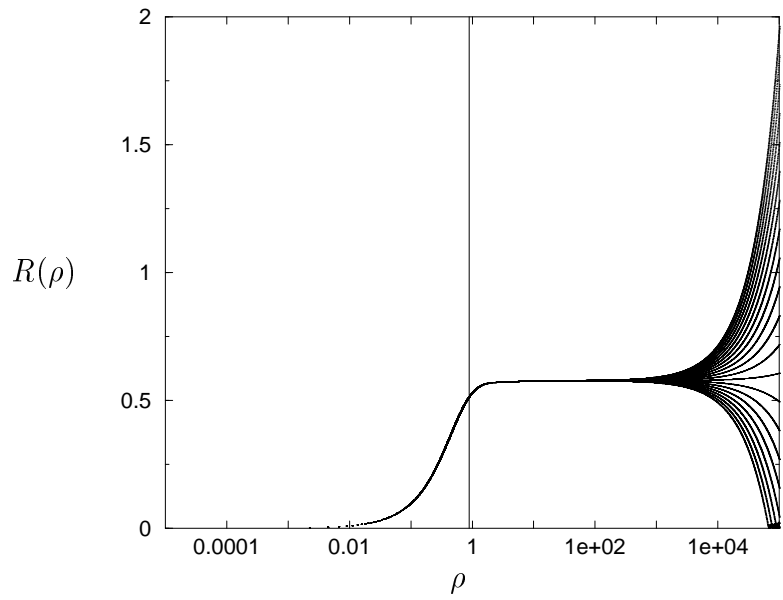
\begin{figure}[h]
\begin{center}
\end{center}
\caption{The area function $R(\rho)$ for the first excitation for $\beta$
near $\beta_{crit}(1)$, 
$0.470366 < \beta_{crit}(1) < 0.470373$. The vertical line marks the horizon
at $\rho = 0.88761$.}
\label{fig:R_rho_Beta_crit} 
\end{figure}

To summarize, we get the following qualitative picture of solutions 
in dependence on the coupling constant $\beta$:
\begin{itemize}
\item For small $\beta$, $0 \le \beta \le \beta_{crit}(n)$, the solutions 
  are similar to those of the uncoupled case $\beta=0$, that is: 
  the area of SO(3) orbits is monotonically increasing with $\rho$, beyond
  the horizon the
  solutions persist up to an infinite value of the coordinate time $\rho$.
  Near infinity the geometry 
  asymptotes to the de Sitter geometry, that is $R = O(\rho)$ and
  $Q$ tends to $-\infty$ as $O(\rho^2)$. According to Hayward's definitions
  \cite{Hayward-trapping_horizon} the horizon is an inner past trapping 
  horizon, separating
  the static region from an expanding dynamic region. 
  The field $f$ shows the same qualitative behavior as in the uncoupled
  case $\beta= 0$ (see Figs. \ref{fig:fn_rho_rhoH} and \ref{fig:endens}).
\item

  For $\beta = \beta_{crit}(n)$ the areal radius $R(\rho)$ still increases
for all $\rho > 0$ but this time goes to a constant at infinity, that is
$R'(\infty) = 0$. For even stronger coupling,   
 $\beta_{crit}(n) < \beta < \beta_{*}(n)$,  $R$ develops a maximum
in the dynamic region, i.e. $R'(\rhoE) = 0$ at some finite time 
$\rhoE > \rhoH$ (see Fig. \ref{fig:R_rho_Beta_crit}).  
As was discussed in Sec. \ref{sec:evolution} $R$ then
decreases and goes to zero at some finite coordinate time $\rhoS$
-- which corresponds to the finite proper time 
$\tau_{S} = \int_{\rhoH}^{\rhoS} d\rho /\sqrt{Q(\rho)}$. This causes the
geometry to be singular at $\rhoS$.
The horizon, which again is an inner
past trapping horizon separates the static region from an initially
expanding dynamic region, which reaches its maximal spatial extension
at $\rhoE$ and then recollapses to a singularity at $\rhoS$.
The maximum of the areal radius occurs at earlier and earlier times as 
the coupling constant is increased 
until it merges with the location of the horizon when $\beta = \beta_{*}(n)$.   
\item 
  At $\beta = \beta_*(n)$ the ``maximal two sphere'' coincides with the
  horizon, $\rhoE = \rhoH$. The fact that $R'(\rhoH) = 0$ at this value
  of $\beta$
  may be interpreted as 
  exchanging the inward and outward direction
  at the horizon: for smaller values of $\beta$ the static region was
  {\em surrounded} by the inner past trapping horizon, 
  whereas for larger values of the coupling
  constant, $\beta_*(n) < \beta < \beta_{max}$, where $\rhoE < \rhoH$,
  the static region {\em encloses} the horizon, which becomes now an
  {\em outer future} trapping horizon. Beyond this horizon, the dynamic
  region undergoes complete collapse at $\rho = \rhoS$. 
\end{itemize}

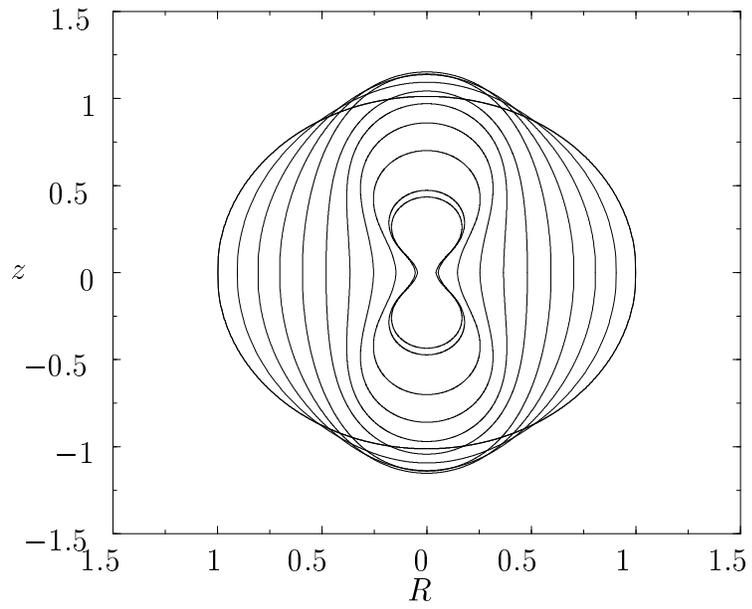
\begin{figure}[h]
\begin{center}
\end{center}
\caption{
The rotational surfaces $z(R)$ correspond to the embedding of a
$t=const$-slice ($\theta = \pi/2$) 
into ${\bf R}^3$.
The upper and lower half part of the diagram resemble the two 
static regions causally separated by the horizon at $z=0$.  
Shown is the geometry of the first excitation
for $\beta$ in the range $0 \le \beta \le 0.913< \beta_{max}$. 
The sphere of unit radius for de Sitter space ($\beta=0$) gets more and more
deformed as $\beta$ increases.
} 
\label{fig:embedding}
\end{figure}

Embedding diagrams of the static regions of the first excitation for several
values of $\beta$ can be found in Fig. \ref{fig:embedding}.
We note, that the ``critical'' values of the coupling constant, 
$\beta_{crit}(n), \beta_*(n)$ and $\beta_{max}(n)$ decrease with the 
excitation number $n$. In Sec. \ref{subsec:staticLambdazero} we will 
give an
argument, that $\beta_{max}(n)$ is a decreasing sequence, which is bounded
from below by the maximal coupling constant $\beta_{max}(\infty) = 1/2$
for the ``singular'' solution (\ref{singularsol}).
 
The solutions described above exist in the presence of a positive
cosmological constant $\Lambda > 0$. As will be described in detail in the
next section, Sec. \ref{subsec:betamax} the limit $\beta \to \beta_{max}$ 
yields regular solutions iff one takes the limit  $\Lambda \to 0$
appropriately as described in Sec. \ref{subsec:betamax}.

\subsection{The Limit $\beta \to \beta_{max}(n)$}\label{subsec:betamax}

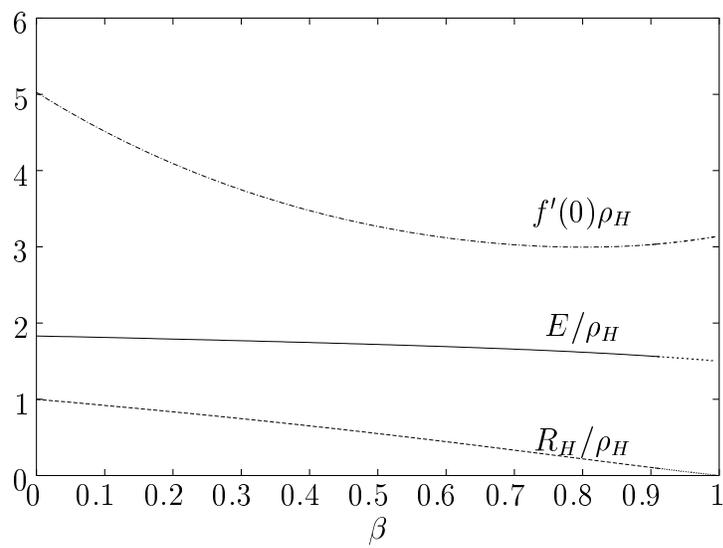
\begin{figure}[h]
\begin{center}
\end{center}
\caption{Some parameters with dimension $length$ measured in units of $\rhoH$.
Except $\rH$ {\it all} of them stay finite in the limit $\beta \to
\beta_{max}$. Moreover in the limit the values tend to the 
corresponding ones of the static Einstein universe, as 
given in table \ref{Lambda0Results}. 
}
\label{paramsscalerhoH}
\end{figure}

Recall from Sec. \ref{subsec:staticEqs} that the cosmological constant
$\Lambda$ sets the length scale in Eqs. (\ref{Q}) - (\ref{constofm}) and
that it can be eliminated from these equations, by introducing
the dimensionless quantities ${\bar \rho} = \sqrt{\Lambda} \rho$ and
${\bar R} = \sqrt{\Lambda} R$. This corresponds to measuring 
all quantities that have
dimension of length, as e.g. the energy $E$, the coordinate distance
of the horizon $\rhoH$ from the origin, the radial geometrical distance 
of the horizon
$\dH$ from the origin, the areal radius $\rH$ of the horizon, 
and $1/f'(0)$,  in units of
$1/\sqrt{\Lambda}$. 
We find that all parameters, that have dimension of length go to zero in the
limit $\beta \to \beta_{max}$ when measured with respect to this length scale.
This indicates that $1/\sqrt{\Lambda}$ is not the appropriate length scale for
taking this limit. We therefore switch to the alternative viewpoint of
$\rhoH$ as our length scale, and we fix $\rhoH = 1$. In this setup
$\Lambda$ depends on $\beta$ and the excitation index $n$ and goes to zero
in the limit $\beta \to \beta_{max}$. The parameters $E$, $\dH$ and
$1/f'(0)$ attain finite values when measured in units of $\rhoH$, whereas
$\rH/\rhoH$ goes to zero. (See Fig. \ref{paramsscalerhoH}).
This strongly suggests, that
there exists a solution with $\beta = \beta_{max}$  
which obeys Eqs. (\ref{Q})-(\ref{constofm})
with $\Lambda = 0$ and has two centers of symmetry.
In particular this means that  
the static region of this solution has
no boundary, since any $t=const$ slice has topology ${\bf S}^3$. 

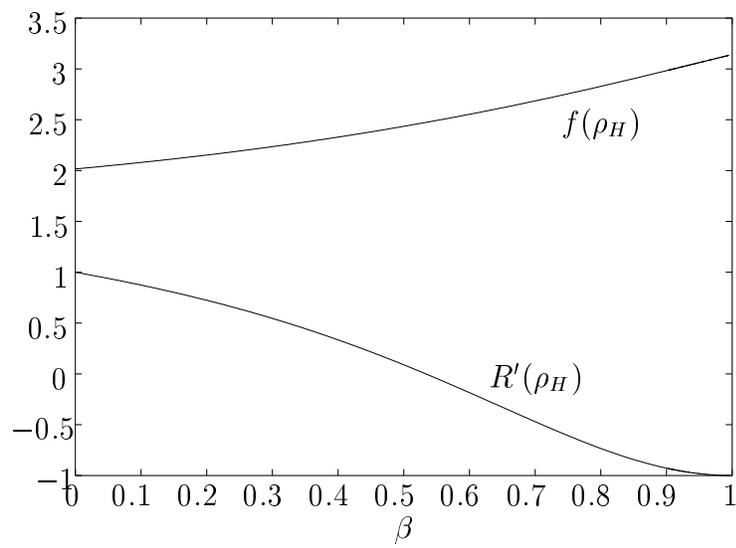
\begin{figure}[h]
\begin{center}    
\end{center} 
\caption{The (dimensionless) parameters $f(\rhoH) \equiv b$ and $R'(\rhoH)$
of the first excitation in dependence of the coupling constant $\beta$.
$f(\rhoH)$ tends to $\pi$ in the limit $\beta \to \beta_{max}$ and
$R'(\rhoH)$ tends to $-1$ in this limit.}
\label{dimlessparams}
\end{figure}

Furthermore, as can be seen from Fig. \ref{dimlessparams}, the dimensionless
parameter $f(\rhoH)$ for the first excitation tends to $\pi$, 
and $R'(\rhoH)$ tends to $-1$ in the limit $\beta \to \beta_{max}$.  
As will be shown in the next section, Sec. \ref{subsec:staticLambdazero}, 
$\Lambda = 0$ implies $Q \equiv 1$.
The limiting solution with $\Lambda = 0$ will therefore
satisfy the regularity conditions (\ref{metricO}) and (\ref{fieldO}) not
only at the axis $\rho = 0$ but also at the second zero of $R$, which
means that such a solution is globally regular with two (regular) centers of
spherical symmetry. 
In fact, 
for the first excitation this limiting solution is just the static Einstein
universe (\ref{EinsteinU}), which can be given in closed form.

These observations allow one to determine the maximal value of the coupling
constant $\beta_{max}(n)$ not as a limiting procedure 
$\beta \to \beta_{max}$, but rather by solving the boundary value problem
Eqs. (\ref{Q}) - (\ref{f}) with $\Lambda =0$ and
with boundary conditions, that correspond to two regular centers 
of symmetry.

\subsection{Globally Static, regular Solutions for $\Lambda =
0$}\label{subsec:staticLambdazero}

For $\Lambda = 0$ Eq. (\ref{Q}) can be solved immediately to give $R^2 Q' =
const$. According to the regularity conditions at the axis (\ref{metricO})
the constant has to vanish, which means that $Q' \equiv 0$ and 
therefore $Q \equiv 1$.
The remaining system of equations is:
\beq\label{feqlambda0}
R'' & = & - \beta R f'^2, \label{Rlambda0}\\
(R^2 f')' & = &\sin (2 f) \label{flambda0}
\eeq
and
\be\label{constofmlabmda0}
2 \beta \sin^{2}f  - \beta R^{2} f'^{2} + R'^{2} - 1 = 0.
\ee
Note, that this system of ODEs is scale invariant, that is any solution
$R(\rho), f(\rho)$ leads via rescaling to the one parameter family of
solutions given by $a R(a \rho), f(a \rho)$. Keeping this in mind, we can
fix the scale arbitrarily, e.g. in setting the first derivative of the field
$f$ equal to one at the origin: $f'(\rho=0) = 1$. Thereby any solution, that
is regular at the origin, is determined {\it entirely } by the value of the
coupling constant $\beta$.

Regularity conditions at the second ``pole'' $R(\rhoP) = 0$ are the same as at
the origin, except that $f$ either tends to $\pi$, if its excitation
number is odd, or to $0$ if it has even excitation number.
This can be inferred from $\pi/2 < f(\rhoH) < \pi$ for $n$ odd and
$0 < f(\rhoH) < \pi/2$ for $n$ even for all $\beta < \beta_{max}$, 
since this is the case for $\beta=0$ and according to (\ref{boundaryH}) no 
crossing of the zero-line or $\pi$-line is allowed.
Note that this corresponds to all odd solutions
having winding number $1$, whereas even solutions are in the topologically
trivial sector.

These regularity conditions together with the invariance of the equations 
under reflection at the location of the maximal two-sphere 
$R'(\rhoE) = 0$, causes globally regular solutions $R(\rho)$ to be 
symmetric around $\rhoE$ whereas $f(\rho) - \pi/2$ is either antisymmetric
for $n$ odd or symmetric
for $n$ even.

For $f$ symmetric the formal power series expansions of $R(\rho)$ and
$f(\rho)$ around $\rho = \rhoE$ gives
\beq\label{fsymm}
R(\rho)& = & R(\rhoE) + O((\rho - \rhoE)^4),  \nonumber\\
f(\rho) & =& \arcsin \sqrt{1/2 \beta} + \frac {2 \sqrt{ 1- 1/2
\beta}}{R(\rhoE)^2 \sqrt{2\beta}} \frac{(\rho -\rhoE)^2}{2!} + O((\rho
-\rhoE)^4),
\eeq
and for $f- \pi/2$ antisymmetric we get  
\beq\label{fantisymm}
R(\rho)& = & \frac{2 \beta - 1}{\beta f'(\rhoE)^2} - (2\beta -1) 
     \frac{(\rho - \rhoE)^2}{2!} + O((\rho - \rhoE)^4),  \nonumber\\
f(\rho) & =& \frac{\pi}{2} + f'(\rhoE)(\rho -\rhoE) + O((\rho-\rhoE)^3).
\eeq

In order to solve the system (\ref{Rlambda0}), (\ref{flambda0}) 
we again use
the shooting and matching method on the interval [origin, $\rhoE$] using the
above Taylor series expansions to determine the boundary conditions at
$\rho = \rhoE$. Shooting parameters are now $\rhoE, f'(\rhoE)$ and $\beta$
for odd solutions and $\rhoE, R(\rhoE)$ and $\beta$ for even solutions.
The results are displayed in Table \ref{Lambda0Results}.

It is clear from (\ref{fsymm}) and (\ref{fantisymm}), 
that regular solutions for 
$\Lambda = 0$ can only exist if $\beta > 1/2$. Assuming now, that our
numerical observations concerning the first few excitations extend to
higher excitations, we give the following argument:
Since
every "branch" of the "$\Lambda > 0$ solutions" persists up to a maximal
value of beta, which can be computed by solving the boundary value problem
(\ref{feqlambda0}) together with regularity conditions at the 
two "poles" -- which implies $ \beta > 1/2 $ -- and since we know, 
that in the limit $\beta \to 0$ there
exists an infinite number of excitations \cite{PCAichCL},
we conclude that this whole family of solutions with $\Lambda > 0$
persists up to some maximal value $\beta_{max}(n)$, which is {\em greater} 
than $1/2$.
In other words, for any $\beta < 1/2 $ there exists a countably infinite
family of solutions with $\Lambda > 0$, whereas for $\beta > 1/2$ our
numerical analysis shows, that only a finite number of solutions exists.
(See Table \ref{Lambda0Results}).

\section{Discussion and Outlook}\label{sec:discussion}

We have shown numerically that the SU(2)-$\sigma$-model coupled to gravity
 with a positive cosmological constant admits a discrete one-parameter family
 of static spherically symmetric regular solutions.
 These solitonic solutions are characterized by an integer excitation
 number $n$.
 A given excitation will only exist up to a critical value of the
 coupling constant $\beta$; the higher $n$, the lower the corresponding
 critical value.
 Our calculations indicate that the infinite tower of solitons present on
 a de Sitter background persists at least up to a value of 
$\beta = 1/2$. Thus 
 there exists a $\beta \ge 1/2$ beyond which the number of excitations 
is finite and decreases with the strength of the coupling.
 As mentioned, qualitatively the $\sigma$-model under consideration 
shows striking
 similarities to the EYM system as studied in detail by Volkov et.al.
 The main difference being that the static solutions to the EYM-system 
 depend
 on the value of the cosmological constant while in our case $\Lambda$ 
 scales
 out from the equations and $\beta$ plays the role of a 
``bifurcation'' parameter.
Another difference concerns the globally regular static solutions with compact 
spatial slices. For the EYM system these appear for definite values of 
$\Lambda(n)$
while for the $\sigma$-model the corresponding solutions exist only in the 
(singular) limit as $\Lambda$ goes to zero and definite values of 
$\beta$. Thus in 
our case there are closed static universes with vanishing cosmological 
constant, the lowest excitation being the static Einstein cosmos. 
This is possible because in this case the stress-energy tensor of the
$\sigma$-field is of the form of a perfect fluid with the equation of state 
$p = -\mu/3$.
Another interesting aspect is the geometry of a given excitation as a function
 of the coupling strength: the static region is always surrounded by a Killing
 horizon separating the static from a dynamical region, which for small 
couplings becomes asymptotically de Sitter. As the coupling is increased the 
two-spheres of symmetry beyond the horizon are first past and then become 
future trapped and a cosmological singularity develops. Finally, for even 
stronger couplings, again the region beyond the horizon collapses, but within 
the static region the in- and outgoing directions (as defined by the sign of 
the expansion for null geodesics) interchange.

An important question to be answered is whether these solitons are stable
 under small radially symmetric time dependent perturbations.
 In a forthcoming publication we intend to present a detailed stability
 analysis. We will show that for $\Lambda > 0$ all excitations are unstable 
with their number 
of unstable modes increasing with $n$. This was to be expected at least for 
small coupling. The lowest excitation thus has a single unstable mode and it 
is known, from other models, that such a solution can play the role of a 
critical solution in a full dynamical treatment of spherically symmetric
collapse.

\acknowledgements
We thank J. Thornburg and M. P\"urrer for computer assistance 
and especially P. Bizon for helpful comments and his interest in
this work. This research was supported in part by FWF as project 
no. P12754-PHY.



\begin{table}[hbt]
\begin{center}

\begin{tabular}{|c|c|c|c|c|}
\hline
$n$ &$\beta_{max}$& $\rhoP = \dP$ & $E/4\pi \gamma$ & $E/4 \pi \gamma \dP$
                                                                \\
\hline
1   &      1      &    $\pi$     & $3 \pi/2$          &   $ 3/2 $    \\
2   &  0.74255    &  6.74225     &  11.78039          &   1.74724   \\
3   &  0.64931    & 12.10140     &  22.43662          &   1.85405   \\
\hline
\end{tabular}
\end{center}
\caption{Results for the first three excitations for $\Lambda = 0$. Since $Q
\equiv 1$
the coordinate distance $\rhoP$ of the two regular ``poles'' equals the radial
geometrical distance $\dP$. The energy density $\rhoP$ and energy $E$ are
 given in units where
$f'(0) = 1$. The ratio $E/\dP$ can be 
compared to the results for solutions
with $\Lambda > 0$ and represents the limit $\beta \to \beta_{max}$ for those
solutions.}
\label{Lambda0Results} 
\end{table}



%
%

%

\end{document}